\documentstyle[12pt]{article}
\renewcommand{\section}[1]{\refstepcounter{section}
\vspace{24pt}\noindent{\bf\arabic{section}.\quad #1}
\vspace*{12pt}}
\newcommand{\ulsect}[1]{\vspace{18pt}\noindent{\bf #1}
\vspace*{12pt}}

\begin{document}
\begin{flushright} KSUCNR-002-94 \\
February 1994\\
\end{flushright}
\vspace*{10mm}
\begin{center}
{\bf Hadron widths in mixed-phase matter}\\[10mm]
David Seibert$^{a*}$ and Che Ming Ko$^b$\\[5mm]
$^a$Physics Department, Kent State University, Kent, OH 44242\\
$^b$Cyclotron Institute and Physics Department,\\
Texas A\&M University, College Station, Texas 77843\\[10mm]
{\bf Abstract}\\
\end{center}
\hspace*{12pt}
We derive classically an expression for a hadron width in a
two-phase region of hadron gas and quark-gluon plasma (QGP). The
presence of QGP gives hadrons larger widths than they would
have in a pure hadron gas.  We find that the $\phi$ width observed
in a central Au+Au collision at $\sqrt{s}=200$ GeV/nucleon
is a few MeV greater than the width in a pure hadron gas.  The
part of observed hadron widths due to QGP is approximately
proportional to $(dN/dy)^{-1/3}$.\\
\vfill
\begin{center}
{\em Submitted to Physical Review Letters}
\end{center}
\vfill
\vspace*{10mm}
\footnoterule
\vspace*{3pt}
$^*$Internet: seibert@scorpio.kent.edu.\\
\newpage\setcounter{page}{1} \pagestyle{plain}
     \setlength{\parindent}{12pt}

There is much interest in the physics of light vector mesons in
ultrarelativistic nuclear collisions [\ref{r1}-\ref{rphi2}].
Experimental studies of the $\rho$ may give a measure of the
transition temperature to quark-gluon plasma (QGP) [\ref{rrho}],
$T_c$, while studies of the $\phi$ can also be used to determine
$T_c$ [\ref{rphi1}] as well as the duration of the transition and
the temperature range over which the transition takes place
[\ref{rphi2}].  However, no previous studies have considered the
effects of QGP on hadron properties observed in ultrarelativistic
nuclear collisions.  These effects can be large, as much of the
hadron signals come from the period during which hadronic matter
and QGP coexist, and hadron properties in QGP are considerably
different from those in hadronic matter.

In this Letter, we derive an expression for a hadron width in
a ``static'' mixed-phase region.  This is a purely classical
derivation, which doesn't include any quantum-mechanical effects.
We then use this derivation to estimate the change in the observed
$\phi$ width in a central Au+Au collision at $\sqrt{s}=200$
GeV/nucleon, and to discuss the effect of varying projectile
and/or target size and collision energy on observed hadron widths.
We use standard high energy conventions, $\hbar=c=k_B=1$,
throughout the letter.

The hadron width in the mixed phase is
\begin{equation}
\Gamma^{\mbox{m}} \, = \, \frac {\displaystyle R^{\mbox{m}}}
{\displaystyle n^{\mbox{m}}},
\end{equation}
where $R^{\mbox{m}}$ and $n^{\mbox{m}}$ are respectively the hadron
decay rate per unit four-volume and density per unit volume in the
mixed phase.  Assuming that hadrons decay but are not created in the
QGP, the decay rate in the QGP is equal to $Area \, \times \,
Flux Density \, \times \, (1-SP)$, where $Area$ is the area
of the hadronic matter -- QGP interface, $Flux Density$ is the
rate per unit area per unit time for hadrons to cross into the QGP,
and $SP$ is the survival probability (the probability that a given
hadron survives a passage through the QGP).  For simplicity, we treat
the mixed-phase matter as a gas of spherical QGP droplets in hadronic
matter; this is reasonable for high energy collisions, as experimental
data is always taken over the whole mixed-phase period, and the
four-volume of QGP is much less than the four-volume of hadronic
matter.

The first two quantities can just be written down:
\begin{eqnarray}
Area &=& \int_0^{\infty} dr \, {\cal P}(r) \, 4\pi r^2, \\
Flux Density &=& \frac {\sigma} {(2\pi)^2} \int_0^{\infty} dk \,
\frac {k^3} {E} \, f_B(E) \int_0^1 dx \, x.
\end{eqnarray}
Here ${\cal P}(r) dr$ is the probability of
finding a QGP droplet with radius between $r$ and $r+dr$, $\sigma$ is
the number of internal degrees of freedom of the hadron, $k$ and
$E=(k^2+m^2)^{1/2}$ are the hadron momentum and energy respectively,
$m$ is the hadron mass, $f_B(E)$ is the
Boltzmann distribution, and the hadron's angle of incidence to the
interface is $\theta = \cos^{-1} x$.  The hadron's velocity is $k/E$,
and the distance the hadron must go to cross the QGP is $2rx$, so the
survival probability is
\begin{equation}
SP = \exp \left[ - \alpha(k) \, r \, x \right],
\end{equation}
where
\begin{equation}
\alpha(k) = 2 E \gamma^{\mbox{q}}(k) /k,
\end{equation}
and $\gamma^{\mbox{q}}(k)$ is the QGP decay rate for a hadron of
momentum $k$.  We thus obtain
\begin{eqnarray}
R^{\mbox{m}} &=& f n^{\mbox{h}} \Gamma^{\mbox{h}} + \frac {\sigma}
{\pi V } \int_0^{\infty} dr \, r^2 \, {\cal P}(r)
\int_0^{\infty} dk \, \frac {k^3} {E} \, f_B(E) \, \int_0^1 dx \,
x \, \left[ 1 - e^{-\alpha(k) r x} \right], \\
 &=& f n^{\mbox{h}} \Gamma^{\mbox{h}} + \frac {\sigma} {\pi V}
\int_0^{\infty} dr \, r^2 \, {\cal P}(r) \int_0^{\infty} dk \,
\frac {k^3} {E} \, f_B(E) \, g\left[ \alpha(k) r \right],
\end{eqnarray}
where $V$ is the volume of hot matter, $f$ is the fraction of
matter in the hadronic phase,
\begin{equation}
n^{\mbox{h}} = \frac {\sigma} {2\pi^2} \int_0^{\infty} dk \, k^2 \,
f_B(E)
\end{equation}
is the hadron density in hadronic matter, and
\begin{equation}
g(z) = \frac 1 2 - \frac {1} {z^2} \left[ 1 - \left( 1 + z \right)
e^{-z} \right].
\end{equation}

The QGP contribution to the $\phi$ density, $n^{\mbox{m}}$, is equal
to $Area \, \times \, Flux Density \, \times \, Time \, / \, V$,
where $Time$ is the mean time a given hadron spends in the QGP before
either decaying or escaping.  The survival probability after time $t$
is $\exp \left[ - \gamma^{\mbox{q}}(k) t \right]$, so we find
\begin{equation}
Time = \frac {1} {\gamma^{\mbox{q}}(k)}
\left[ 1 - e^{- \alpha(k) r x} \right].
\end{equation}
Combining this with the expressions for $Area$ and $Flux Density$
above, we obtain
\begin{eqnarray}
n^{\mbox{m}} &=& f n^{\mbox{h}} + \frac {\sigma} {\pi V}
\int_0^{\infty} dr \, r^2 \, {\cal P}(r) \int_0^{\infty} dk \,
\frac {k^3} {E} \, f_B(E) \, \int_0^1 dx \, x \,
\frac {1} {\gamma^{\mbox{q}}(k)} \left[ 1 -
e^{- \alpha(k) r x} \right], \\
&=& f n^{\mbox{h}} + \frac {\sigma} {\pi V}
\int_0^{\infty} dr \, r^2 \, {\cal P}(r) \int_0^{\infty} dk \,
\frac {k^3} {E} \, f_B(E) \, \frac {1} {\gamma^{\mbox{q}}(k)} \,
g\left[ \alpha(k) r \right].
\end{eqnarray}

If the decay rate in QGP is small [$\gamma^{\mbox{q}} \ll k/(rE)$
for thermal momenta], we expand the exponentials and find
\begin{eqnarray}
R^{\mbox{m}} &=& f n^{\mbox{h}} \Gamma^{\mbox{h}} + \frac {2 \sigma}
{\pi V} \int_0^{\infty} dr \, r^3 \, {\cal P}(r) \int_0^{\infty} dk \,
k^2 \, f_B(E) \, \gamma^{\mbox{q}}(k), \\
&=& n^{\mbox{h}} \left[ f \Gamma^{\mbox{h}} + (1-f) \Gamma^{\mbox{q}}
\right], \\
n^{\mbox{m}} &=& n^{\mbox{h}},
\end{eqnarray}
where
\begin{equation}
\Gamma^{\mbox{q}} = \frac {\displaystyle \int_0^{\infty} dk \, k^2 \,
f_B(E) \, \gamma^{\mbox{q}}(k)} {\displaystyle \int_0^{\infty} dk \,
k^2 \, f_B(E)}
\end{equation}
is the thermally averaged hadron decay rate in QGP.  In deriving
these results, we used the normalization condition
\begin{equation}
\int_0^{\infty} dr \, {\cal P}(r) \, \frac 4 3 \pi r^3 = (1-f) \, V.
\end{equation}
Dividing the decay rate by the density, we obtain the simple expression
\begin{equation}
\Gamma^{\mbox{m}} = f \Gamma^{\mbox{h}} + (1-f) \Gamma^{\mbox{q}}.
\end{equation}

If the decay rate is large, the expression for the width is not so
simple.  In the limit $\gamma^{\mbox{q}}(k) \rightarrow \infty$, we
obtain
\begin{eqnarray}
R^{\mbox{m}} &=& f n^{\mbox{h}} \Gamma^{\mbox{h}} + \frac {\sigma}
{2\pi V} \int_0^{\infty} dr \, r^2 \, {\cal P}(r)
\int_0^{\infty} dk \, \frac {k^3} {E} \, f_B(E), \\
n^{\mbox{m}} &=& f n^{\mbox{h}}, \\
\Gamma^{\mbox{m}} &=& \Gamma^{\mbox{h}}
+ \frac {\sigma} {2\pi f V n^{\mbox{h}}}
\int_0^{\infty} dr \, r^2 \, {\cal P}(r)
\int_0^{\infty} dk \, \frac {k^3} {E} \, f_B(E). \label{eginf}
\end{eqnarray}
Note that, in this case, the width depends on the details of the
mixed phase structure through ${\cal P}(r)$, and not just on the
fraction of QGP that is present.

Unfortunately, the only way to calculate ${\cal P}(r)$
is to do a three-dimensional simulation with both nucleation and
growth of droplets of hadronic phase in the QGP.  This is because
${\cal P}(r)$ is really an inherently dynamical quantity, and
goes to a $\delta$-function in a static mixed phase.  If you
begin with several droplets of one phase, the volume energy is
independent of the number of droplets since total volume in the
phase is conserved, but the surface energy grows with the number of
droplets, so the free energy is minimized when all of the phase is
in one large droplet.  This can be seen from droplet nucleation
and growth calculations [\ref{rkcs}] -- if the droplet radius is
larger than some critical radius, the droplet grows (forever, or
until it runs out of matter), so any droplet distribution other
than a single droplet is metastable.

If we specify the volume, $V$, then the droplet distribution is
known:
\begin{eqnarray}
{\cal P}(r) &=& \delta (r-r_d), \\
\frac {4\pi} {3} r_d^3 &=& (1-f) V. \label{erdt}
\end{eqnarray}
We then evaluate the integrals over $r$ to get
\begin{eqnarray}
R^{\mbox{m}} &=& f n^{\mbox{h}} \Gamma^{\mbox{h}} +
\frac {3 \sigma (1-f)} {4 \pi^2 r_d} \int_0^{\infty} dk \,
\frac {k^3} {E} \, f_B(E) \, g\left[ \alpha(k) r_d \right], \\
n^{\mbox{m}} &=& f n^{\mbox{h}} +
\frac {3 \sigma (1-f)} {4 \pi^2 r_d} \int_0^{\infty} dk \,
\frac {k^3} {E \gamma^{\mbox{q}}(k)} \, f_B(E) \,
g\left[ \alpha(k) r_d \right].
\end{eqnarray}

We now move to the specific case of the observed change in the
$\phi$ width in an ultrarelativistic nuclear collision.  The $\phi$
decay rate in the QGP, $\gamma^{\mbox{q}}(k)$, can be estimated by
modeling the $\phi$ propagating through the QGP as an $s\overline{s}$
pair.  We thus obtain
\begin{equation}
\gamma^{\mbox{q}} \approx 2 \gamma_s,
\end{equation}
where $\gamma_s$ is the width of a thermal $s$ (or $\overline{s}$),
as we destroy the $\phi$ if either the $s$ or $\overline{s}$ is
scattered in the QGP.

Treating the propagating $s$ and $\overline{s}$ as light quarks with
thermal momenta, we find from Pisarski [\ref{rpis}] that
\begin{equation}
\gamma^{\mbox{q}} = 2 \alpha_S C_f T_c
\ln \left[ \frac {1} {\alpha_S} \right],
\end{equation}
independent of $k$.
Here $\alpha_S=g^2/4\pi$, $g$ is the strong coupling constant,\break
$C_f=(N^2-1)/(2N)$ is the Casimir factor for the fermion
representation, $N$ is the number of colors, and the argument of the
logarithm is an approximation to what one would get with a magnetic
mass cutoff.  Note that $\gamma^{\mbox{q}}(k)$ is four times the
parton damping rate [the damping rate is for the wavefunction, but
you square the wavefunction to get the probability density so the
parton decay rate is twice the damping rate, and the $\phi$ decay
rate is twice the parton decay rate].  With $N=3$, $C_f=4/3$, and
$g \simeq 2$, we find
\begin{equation}
\gamma^{\mbox{q}} \simeq T_c.
\end{equation}

To put this in a simple one-dimensional (no transverse expansion),
boost-invariant, hydrodynamic collision simulation~[\ref{rbj}], we
assume that we have an approximately massless gas so that $\tau T^3$
is conserved outside the mixed phase region, where $\tau$ is proper
time.  In the mixed phase region, the requirement of entropy
conservation relates $f$ and $\tau$:
\begin{equation}
f(\tau) = \frac {r} {r-1} \left( \frac {\tau-\tau_q} {\tau} \right),
\label{eft} \end{equation}
where $r$ is the ratio of number of degrees of freedom in the QGP
and hadronic phases, and $\tau_q$ is the proper time at which the
period of two-phase coexistence begins.

During the early stages of the evolution, the behavior of
${\cal P}(r)$ is very complicated.  However, the volume of hadronic
matter is greatest near the end of the mixed-phase period, so most
of the observed hadrons occur during the end of this stage.  Late in
the mixed phase, we can try to guess the droplet distribution.
Because the droplets with similar velocities tend to merge, we
assume that (i) there is only one droplet in any transverse slice,
and (ii) the droplets are distributed uniformly in rapidity with
distance $y_d$ between droplets.  Thus, we find
\begin{equation}
V(\tau) = A y_d \tau, \label{eVt}
\end{equation}
where $A$ is the cross-sectional area of the hot matter.

Combining eqs.~(\ref{erdt}), (\ref{eft}), and (\ref{eVt}), we obtain
\begin{equation}
r_d(\tau) = \left[ \frac {3 A y_d} {4 \pi (r-1)}
\left( r\tau_q - \tau \right) \right]^{1/3}.
\end{equation}
The observed width is then
\begin{equation}
\Gamma^{\mbox{m}} = \frac
{\displaystyle r (r-1)^2 \tau_q^2 n^{\mbox{h}} \Gamma^{\mbox{h}}
+ \frac {3 \sigma} {2 \pi^2} \int_{\tau_q}^{r\tau_q} d\tau \,
\frac {r\tau_q-\tau} {r_d(\tau)} \int_0^{\infty} dk \,
\frac {k^3} {E} \, f_B(E) \, g\left[ \alpha(k) r_d(\tau) \right]}
{\displaystyle r (r-1)^2 \tau_q^2 n^{\mbox{h}} +
\frac {3 \sigma} {2 \pi^2 \gamma^{\mbox{q}}}
\int_{\tau_q}^{r\tau_q} d\tau \, \frac {r\tau_q-\tau} {r_d(\tau)}
\int_0^{\infty} dk \, \frac {k^3} {E} \, f_B(E) \,
g\left[ \alpha(k) r_d(\tau) \right]}.
\end{equation}
For a central Au+Au collision at $\sqrt{s}=200$ GeV/nucleon, we
take $A=150$ fm$^2$, $r=10$, and $\tau_q=8$ (16) fm/$c$ for
$T_c=190$ (150) MeV.  For $T_c=150$ MeV, we find
$\Gamma^{\mbox{m}}-\Gamma^{\mbox{h}} = 1.1$ MeV with $y_d=1$ and
2.3 MeV with $y_d=0.1$, while for $T_c=190$ MeV we find 1.5 and
3.1 MeV respectively.  These values are smaller than the width in
a hadronic gas (about 6 MeV at $T=150$ MeV and 9 at $T=190$ MeV
[\ref{rphi2}]), but are still large enough to be significant.

We obtain a second estimate of $\gamma^{\mbox{q}}(k)$ from the
damping rate in the heavy quark limit [\ref{rpis}],
\begin{equation}
\gamma^{\mbox{q}}(k) = 2 \alpha_S C_f T_c \left\{ 1 +
\frac {k} {2E} \ln \left[ \frac {4 \pi^2 (N + N_f/2) k^3}
{3 C_f^2 \alpha_S E^3} \right] \right\},
\end{equation}
where $N_f$ is the number of light fermion species.  Taking $N=3$,
$C_f=4/3$, $N_f=3$ ($u$, $d$, $s$) and $g=2$,
\begin{equation}
\gamma^{\mbox{q}}(k) = \frac {8} {3\pi} T_c \left[ 1 +
\frac {3k} {2E} \ln \left( \frac {3 \pi k} {2 E} \right) \right].
\end{equation}
With this new expression for $\gamma^{\mbox{q}}(k)$, we find
$\Gamma^{\mbox{m}}-\Gamma^{\mbox{h}} = 1.1$ MeV with $y_d=1$ and
2.4 MeV with $y_d=0.1$ for $T_c=150$ MeV, while we find 1.5 and
3.3 MeV respectively for $T_c=190$ MeV.  Thus, our results are
insensitive to our assumptions about $\gamma^{\mbox{q}}$.

In the limit $\gamma^{\mbox{q}}(k) \rightarrow \infty$, we find
\begin{eqnarray}
\Gamma^{\mbox{m}}-\Gamma^{\mbox{h}} &=&
\frac {9 \, \overline{v}(T_c)} {5 \, r} \left[ \frac {\pi}
{6 \, A \, y_d \, \tau_q} \right]^{1/3}, \label{eginf2} \\
\overline{v}(T_c) &=& \frac
{\displaystyle \int_0^{\infty} dk \, \frac {k^3} {E} \, f_B(E)}
{\displaystyle \int_0^{\infty} dk \, k^2 \, f_B(E)}.
\end{eqnarray}
For $T_c=150$ MeV, we obtain
$\Gamma^{\mbox{m}}-\Gamma^{\mbox{h}}=1.2$  MeV with $y_d=1$ and
2.5 MeV with $y_d=0.1$, while for $T_c=190$ MeV we obtain 1.6
and 3.4 MeV respectively.  We also calculate $\Gamma^{\mbox{m}}$
as a function of total rapidity density (charged plus neutral),
$dN/dy$, using the relation
\begin{equation}
3.6 \, dN/dy = \frac {74 \pi^2} {45} \, A \, \tau_q \, T_c^3
\end{equation}
(from entropy conservation) to obtain
\begin{equation}
\Gamma^{\mbox{m}}-\Gamma^{\mbox{h}} =
\frac {\pi \, T_c \, \overline{v}(T_c)} {5 \, r}
\left[ \frac {111} {2 \, (dN/dy) \, y_d} \right]^{1/3}.
\label{eginf0} \end{equation}
This last form is probably the most useful for experimenters, as
$dN/dy$ is readily measured, unlike $A$ and $\tau_q$.

Note that we do not include damping due to Debye screening, as the
$\phi$ is already formed, unlike the case of J/$\psi$ suppression
[\ref{rdeb}].
As long as no constituents of the $\phi$ scatter, it should emerge
from the QGP with its wavefunction intact.  We also neglect the
possibility of reflection from the hadronic matter -- QGP interface.
However, if we take the limit $\gamma^{\mbox{q}} \rightarrow \infty$
[eq.~(\ref{eginf2})], our results are changed very little.  Thus,
adding effects that increase $\gamma^{\mbox{q}}$ does not produce
a large change in the $\phi$ width.

The main effect of changing the projectile or target is to change
$A$.  As $\Gamma^{\mbox{m}}$ only depends on $A$ through $r_d$,
and $r_d$ increases with increasing $A$, changing $A$ is equivalent
to changing $r_d$.  For large $\gamma^{\mbox{q}}(k)$, which is the
case for essentially any hadron in QGP, $\Gamma^{\mbox{m}}$ decreases
monotonically with increasing $r_d$ and thus with increasing $A$ or
$y_d$.  Therefore, $\Gamma^{\mbox{m}}$ will be greater in collisions
of smaller projectiles and/or targets.

Similarly, if collision energy is lowered then $\tau_q$ is
reduced.  This reduces $r_d$, thus increasing $R^{\mbox{q}}$.  All
other changes in $\Gamma^{\mbox{m}}$ can be absorbed in the
normalization of the integrals, so lowering the collision energy
will increases $\Gamma^{\mbox{m}}$.  Decreasing the projectile
and/or target size also tends to decrease $\tau_q$ slightly, and
this will increase hadron widths.  Finally, raising $T_c$ reduces
$\tau_q$, which is why $\Gamma^{\mbox{m}}-\Gamma^{\mbox{h}}$ is
30-40\% larger at $T_c=190$ MeV than at $T_c=150$ MeV.  This
qualitative behavior -- the increase of
$\Gamma^{\mbox{m}}-\Gamma^{\mbox{h}}$ with decreasing
projectile and/or target size, decreasing collision energy and
increasing $T_c$ -- will hold for any hadron.

The portion of the hadron width due to QGP can be observed
experimentally using satellite vector meson peaks in the dilepton
spectrum, which are a signal of a strong first-order hadronic phase
transition [\ref{rphi1}].  These satellite peaks come from mesons that
decay in the mixed phase, at fixed temperature, so the meson width in a
hadron gas is constant.  The increase in the width due to QGP droplets
can be observed by comparing the widths in events with different
values of $dN/dy$, as long as conditions are such that thermal
equilibration should occur before the region of mixed-phase matter is
formed.  The portion of the width due to the QGP droplets should scale
approximately as $(dN/dy)^{-1/3}$, following eq.~(\ref{eginf0}),
since $T_c$ and $r$ do not vary between events and $y_d$ probably
doesn't vary much.

Note that this increased width is not an unambiguous signal for the
presence of QGP.  Suppose that the nature of the high-temperature phase
is different (perhaps chiral symmetry is restored without deconfinement),
but the hadron lifetimes are still short in this phase.  In that case,
the calculation would proceed just as above, and the effect of droplets
of high-temperature phase on the hadron widths would be essentially
unchanged.  Thus, the observation of this increased width is a signal
only for short hadron lifetimes in the high-temperature phase.

\bigskip

We thank George Fai and Declan Keane for helpful discussions.  The
work of D.S. was supported in part by the U.S. Department of Energy
under Grant No.\ DOE/DE-FG02-86ER-40251. The work of C.M.K was
supported in part by the National Science Foundation under Grant No.\
PHY-9212209 and the Welch Foundation under Grant No.\ A-1110.

\vfill \eject

\ulsect{References}

\begin{list}{\arabic{enumi}.\hfill}{\setlength{\topsep}{0pt}
\setlength{\partopsep}{0pt} \setlength{\itemsep}{0pt}
\setlength{\parsep}{0pt} \setlength{\leftmargin}{\labelwidth}
\setlength{\rightmargin}{0pt} \setlength{\listparindent}{0pt}
\setlength{\itemindent}{0pt} \setlength{\labelsep}{0pt}
\usecounter{enumi}}

\item \label{r1} E. Shuryak, Nucl.\ Phys.\ {\bf A544}, 65c (1992);
T. Hatsuda, Nucl.\ Phys.\ {\bf A544}, 27c (1992);
U. Heinz and K.S. Lee, Nucl.\ Phys.\ {\bf A544}, 503c (1992);
C. Sch\"uren {\it et al.}, Nucl.\ Phys.\ {\bf A565}, 687 (1993);
S. Kim and D.K. Sinclair, Phys.\ Rev.\ D {\bf 48}, 4408 (1993);
C. Bernard {\it et al.}, Phys.\ Rev.\ D {\bf 48}, 4419 (1993);
C.R. Allton {\it et al.}, Phys.\ Rev.\ D {\bf 49}, 474 (1994).

\item \label{rrho} D. Seibert, Phys.\ Rev.\ Lett.\ {\bf 68} (1992) 1476;
D. Seibert, V.K. Mishra, and G. Fai, Phys.\ Rev.\ C {\bf 46} (1992) 330.

\item \label{rphi1}  M. Asakawa and C.M. Ko, Phys.\ Lett.\ B
(in press); C.M. Ko and M. Asakawa, Nucl.\ Phys.\ {\bf A566}, 447c
(1994).

\item \label{rphi2} C.M. Ko and D. Seibert, Phys.\ Rev.\ C (in press).

\item \label{rkcs} L.P. Csernai and J.I. Kapusta, Phys.\ Rev.\ Lett.\
{\bf 69}, 737 (1992); L.P. Csernai, J.I. Kapusta, Gy. Kluge, and E.E.
Zabrodin, Z. Phys.\ C {\bf 58}, 453 (1993).

\item \label{rpis} R. Pisarski, Phys.\ Rev.\ D {\bf 47}, 5589 (1993).

\item \label{rbj} J.D. Bjorken, Phys.\ Rev.\ D {\bf 27}, 140 (1983).

\item \label{rdeb}
T. Matsui and H. Satz, Phys.\ Lett.\ B {\bf 178}, 416 (1986);
A. Capella {\it et al.}, Phys.\ Lett.\ B {\bf 206}, 354 (1988);
J. Ftanik, P. Lichard, and J. Pisut, Phys.\ Lett.\ B {\bf 207}, 194
(1988);
S. Gavin, M. Gyulassy, and A. Jackson, Phys.\ Lett.\ B {\bf 207},
257 (1988);
R. Vogt {\it et al.}, Phys.\ Lett.\ B {\bf 207}, 263 (1988);
J.P. Blaizot and J.Y. Ollitrault, Phys.\ Lett.\ B {\bf 217}, 386 (1989).

\end{list}

\vfill\eject

\end{document}